\documentclass[aps,12pt,preprintnumbers,nofootinbib,superscriptaddress]{revtex4}
\usepackage[letterpaper]{geometry}                % See geometry.pdf to learn the layout options. There are lots.
\usepackage{graphicx}
\usepackage{amssymb,amsmath}
\usepackage{feynmf}
\DeclareMathOperator{\tr}{tr}

\def\nn{\nonumber \\ }

\def\vev#1{\left\langle #1 \right\rangle}
\def\tr{\text{Tr}\,}
\begin{document}

\newcount\hour \newcount\minute
\hour=\time \divide \hour by 60
\minute=\time
\count99=\hour \multiply \count99 by -60 \advance \minute by \count99
\newcommand{\mydate}{\ \today \ - \number\hour :00}

\preprint{CALT 68-2652}

\title{Color Octet Scalar Production at the LHC}

\author{Moira I. Gresham}
\email[]{moira@theory.caltech.edu}
\affiliation{California Institute of Technology, Pasadena, CA 91125}

\author{Mark B. Wise}
\email[]{wise@theory.caltech.edu}
\affiliation{California Institute of Technology, Pasadena, CA 91125}

\date{\today}                                           

\begin{abstract}

New physics at the weak scale that can couple to quarks typically gives rise to unacceptably large flavor changing neutral currents.  An attractive way to avoid this problem is to impose the principal of minimal flavor violation (MFV). Recently it was noted that in MFV only scalars with the same gauge quantum numbers as the standard model Higgs doublet or color octet scalars with the same weak quantum numbers as the Higgs doublet can couple to quarks. In this paper we compute the one-loop rate for production of a single color octet scalar through gluon fusion at the LHC, which can become greater than the tree level pair production rate for octet scalar masses around a TeV. We also calculate the precision electroweak constraint from $Z \rightarrow \bar b b$; this constraint on color octet mass and Yukawa coupling affects the allowed range for single octet scalar production through gluon fusion.

\end{abstract}

\maketitle
\newpage
\section{Introduction}

The standard model for strong, weak, and electrodynamics has been very successful. Nonetheless, most physicists expect that the next generation of very high energy accelerator experiments that are able to produce resonances with masses around the ${\rm TeV}$ scale will find physics beyond what is in the minimal standard model. There are two reasons for this. Firstly, the mechanism for weak symmetry breaking in the minimal standard model, a single scalar doublet, is the simplest but has no direct confirmation from experiment. Secondly, an awkward fine-tuning of parameters must be made, order by order in perturbation theory, to keep the physical mass of the Higgs scalar very light compared to the ultraviolet cutoff. This is called the hierarchy problem and is the motivation for most of the proposed extensions of the minimal standard model with new physics at the weak scale.

Models with new degrees of freedom at the weak scale typically give rise to unacceptably large flavor changing neutral currents (FCNC) if the new physics can couple at tree level to the quarks and the most general couplings are allowed. The minimal standard model does not have such a problem because of the GIM mechanism. More generally, the large FCNC problem does not arise if the $SU(3)_{Q_L}\times SU(3)_{U_R}\times SU(3)_{D_R}$ quark flavor symmetry is only broken by a single pair of Higgs Yukawa coupling matrices $g^U$ and $g^D$. This way of suppressing FCNC is called minimal flavor violation (MFV)~\cite{mfv}. In MFV, only scalars with the same gauge quantum numbers as the Higgs doublet or color octet scalars with the same weak quantum numbers of the Higgs doublet can Yukawa couple to the quarks \cite{wise2006a}. Therefore it is of interest to study the properties of such scalars. Models with several color singlet, weak doublet scalars have been extensively studied. In~\cite{popov}, certain decays of color octet scalars were studied in the context of Pati-Salam unification. Here, we continue the phenomenological analysis of color octet scalars begun in \cite{wise2006a}. 

The tree level pair production cross section for charged or neutral color octet scalars was computed in~\cite{wise2006a}. The precision electroweak variables $S$ and $T$ were also computed\footnote{The parameter $U$ is small.} and a number of FCNC processes were considered. Here, we calculate the one-loop production rate for a single neutral octet scalar through gluon fusion, $gg \rightarrow S^0$. We also derive the constraint on the strength of the coupling of color octet scalars to up-type quarks that arises from experimental data on $R_b$. This constraint restricts the magnitude of the the $gg \rightarrow S^0$ cross~section.

\section{The Model}

The standard model quark Yukawa couplings of the quarks to the Higgs doublet  $H$ are 
\begin{eqnarray}
L &=& -g_{ij}^U \bar u_{Ri}  Q_j H - g_{ij}^D \bar{d}_{Ri}  Q_j H^\dagger + \text{h.c.},
\label{yuk}
\end{eqnarray}
where $i$ and $j$ are flavor indices, and gauge indices have been omitted. Repeated flavor indices are summed over. The Yukawa couplings generate the mass matrices
\begin{eqnarray}
M_{ij}^U  = g_{ij}^U \vev{H^2},~~~~  M_{ij}^D = g_{ij}^D \vev{H^2}^\dagger,
\label{mass}
\end{eqnarray}
for the charge $2/3$ and $-1/3$ quarks  when the Higgs field gets a vacuum expectation value $\vev{H^1}=0,~\vev{H^2}=v/\sqrt{2}$. In the minimal standard model the only couplings that violate the $SU(3)_{Q_L}\times SU(3)_{U_R}\times SU(3)_{D_R}$ quark flavor symmetry are the Yukawa matrices $g^U$ and $g^D$. We can view the theory as being invariant under the flavor group if the Yukwawa matrices are endowed with the transformation property,
\begin{equation}
g^U \rightarrow V_U g^U V_Q^{\dagger} \qquad g^D\rightarrow V_D g^D V_Q^{\dagger},
\end{equation}
where $V_U$ is an element of  $SU(3)_{U_R}$, $V_D$ is an element of  $SU(3)_{D_R}$ and $V_Q$ is an element of  $SU(3)_{Q_L}$.

In this paper we add to the minimal standard model a single weak doublet of color octet scalars $S^A$. According to MFV, its Yukawa couplings to the quarks are,
\begin{eqnarray}
L &=& - \eta_U { \bar g}_{ij}^U \bar u_{Ri}T^A  Q_j S^A- \eta_D {\bar g}_{ij}^D  \bar d_{Ri} T^A Q_j S^{A\,\dagger}+\text{h.c.}~,%\nn
\label{yuk8}
\end{eqnarray}
where the matrices ${ \bar g}^U$ and ${\bar g}^D$ also transform as 
\begin{equation}
{\bar g}^U \rightarrow V_U {\bar g}^U V_Q^{\dagger} \qquad {\bar g}^D\rightarrow V_D {\bar g}^D V_Q^{\dagger},
\end{equation}
and are composed from $g^U$ and $g^D$. So,
\begin{equation}
\label{series}
{\bar g}^U= g^U+\epsilon_1^U g^U (g^D)^{\dagger}g^D + ...
\end{equation} 
and
\begin{equation}
{\bar g}^D= g^D+\epsilon_1^D g^D (g^U)^{\dagger}g^U + ...~~ .
\end{equation} 
Note that there is a term in the ellipses of Eq.~(\ref{series}) proportional to  $g^U (g^U)^{\dagger}g^U$; however, it does not give rise to flavor changing neutral current effects, so we neglect it. We will assume that the $\epsilon ^{U,D}$ are small and that terms with more powers of the Yukawa couplings are more suppressed and can be neglected.
Diagonalizing the quark mass matrices we find, in the quark mass eigenstate basis, that the couplings of the octet scalars take the form,
\begin{eqnarray}
L&=&-\sqrt{2}\eta_U{\bar u}_{Ri} \frac{m^U_i}{v} \left( \delta_{ij}+2\epsilon_1^UV_{ik}(m^D_k/v)^2 V_{kj}^{\dagger}+...~~ \right) T^A u_{Lj} S^{A0} \nonumber \\
&+&\sqrt{2}\eta_U{\bar u}_{Ri} \frac{m^U_i}{v} V_{ij}\left( 1+2\epsilon_1^U(m^D_j/v)^2 +...~~ \right) T^A d_{Lj} S^{A+} \nonumber \\
&-&\sqrt{2}\eta_D{\bar d}_{Ri} \frac{m^D_i}{v} \left( \delta_{ij}+2\epsilon_1^DV_{ik}^{\dagger}(m^U_k/v)^2 V_{kj}+...~~ \right) T^A d_{Lj} S^{A0\dagger} \nonumber \\
&-&\sqrt{2}\eta_D{\bar d}_{Ri} \frac{m^D_i}{v} \left( 1+2\epsilon_1^D(m^U_j/v)^2 +...~~ \right) V_{ij}^{\dagger}T^A u_{Lj} S^{A\,-} + {\rm h.c.},
\end{eqnarray}
where $V$ is the CKM matrix and $m^U_i$, $m^D_i$ are the charge $2/3$ and charge $-1/3$ quark masses. Since the top quark is by far the heaviest quark we can approximate the above by,
\begin{eqnarray}
L&=&-\sqrt{2}\eta_U{\bar u}_{Ri} \frac{m^U_i}{v} T^A u_{Li} S^{A0} -\sqrt{2}\eta_D{\bar d}_{Ri} \frac{m^D_i}{v} V_{ij}^{\dagger}T^A u_{Lj} S^{A\,-}
+\sqrt{2}\eta_U{\bar u}_{Ri}  \frac{m^U_i}{v} V_{ij}T^A d_{Lj} S^{A+} \nonumber \\
&-&\sqrt{2}\eta_D{\bar d}_{Ri} \frac{m^D_i}{v} \left( \delta_{ij}+2\epsilon_1^DV_{i3}^{\dagger}(m_t/v)^2 V_{3j}+...~~ \right) T^A d_{Lj} S^{A0\dagger}
+ {\rm h.c.}.
\end{eqnarray}
The term proportional to $\epsilon_1^D$ gives rise to a tree level contribution to flavor changing neutral current processes like $ B \bar B $ mixing from $S^0$ exhange. However, the leading contribution to $ B \rightarrow X_S \gamma$ does not involve $\epsilon_1^D$ and experimental data on this process provides an important constraint on $\eta_D$. The parameter $\eta_U$  is constrained from data on the precision electroweak variable $R_b$; this is discussed in the next section.

The most general renormalizable scalar potential is~\cite{wise2006a},

\begin{eqnarray}
V &=& \frac{\lambda}{4}\left(H^{\dagger i} H_i-\frac {v^2}{2}\right)^2 + 2 m_S^2 \tr S^{\dagger i} S_i
 +\lambda_1 H^{\dagger i} H_i \tr S^{\dagger j} S_j 
+ \lambda_2 H^{\dagger i} H_j \tr S^{\dagger j} S_i\nn
&& + \Bigl[ \lambda_3  H^{\dagger i} H^{\dagger j} \tr S_i S_j  + \lambda_4 H^{\dagger i} \tr S^{\dagger j} S_j  S_i 
+ \lambda_5 H^{\dagger i} \tr S^{\dagger j} S_i  S_j + \text{h.c.}\Bigr]\nn
&&+ \lambda_6 \tr S^{\dagger i} S_i S^{\dagger j} S_j 
+ \lambda_7 \tr S^{\dagger i} S_j S^{\dagger j} S_i 
+ \lambda_8 \tr S^{\dagger i} S_i  \tr S^{\dagger j} S_j 
+ \lambda_9 \tr S^{\dagger i} S_j  \tr S^{\dagger j} S_i \nn
&&+\lambda_{10} \tr  S_i   S_j  S^{\dagger i} S^{\dagger j}
+ \lambda_{11} \tr  S_i  S_j  S^{\dagger j} S^{\dagger i} .\nn 
\end{eqnarray}
We have explicitly displayed the $SU(2)$ indices on the Higgs doublet and on the color octet scalars. Traces are over color indices and the notation $S=S^AT^A$ is used, where the $SU(3)$ generators  have their standard normalization: $\tr T^A T^B=\delta^{AB}/2$. The coupling $\lambda_3$ has been made real by a phase rotation of the $S$ fields. With this phase convention the phases of $\eta_{U,D}$ and $\lambda_{4,5}$ represent additional sources of CP violation beyond those in the minimal standard model.

The Higgs vacuum expectation value causes a tree level mass splitting between the octet scalars. It is convenient to decompose the neutral complex octet scalars into two real scalars,
\begin{equation}
S^{A0}={S^{A0}_R+i S^{A0}_I \over {\sqrt 2}}.
\end{equation}
Then the tree level mass spectrum is~\cite{wise2006a},
\begin{eqnarray}
m^2_{S^{\pm}}&=&m_S^2+\lambda_1{ v^2 \over 4} ,\nn
m^2_{S_R^0}&=&m_S^2+\left(\lambda_1+\lambda_2+2\lambda_3\right){v^2 \over 4} ,\nn
m^2_{S_I^0}&=&m_S^2+\left(\lambda_1+\lambda_2-2\lambda_3\right){v^2 \over 4}.
\label{mass}
\end{eqnarray}

In this paper we focus on color octet scalars with masses greater than $500~{\rm GeV}$. The mass splittings are expected to be
small compared with this and so we neglect the mass splittings between the various color octet scalar states for the remainder of this paper. Color octet scalars with masses between $500$GeV and $1$TeV can have a dramatic impact on the rate for Higgs production at the LHC \cite{wise2006a, wise2006b}.

\section{Precision Electroweak Constraints}

In \cite{wise2006a} the values of the oblique parameters $S$ and $T$ that arise in this model were computed. These corrections to the standard model are expressed in terms of parameters in the scalar potential (including the color octet scalar masses). The contribution to the effective Hamiltonian for $b \rightarrow s \gamma$  proportional to $\eta_U \eta_D$ was also considered. Here we complete the analysis of precision electroweak physics in this theory by computing $R_b$, the ratio of the $Z$ width to final hadronic states containing a $b$ and $\bar b$ quark to the total hadronic width. We write the coupling of the $Z$ boson to quarks $q$ as,
\begin{equation}
-{g_2 \over \cos \theta_W}Z^{\mu} {\bar q}\gamma_{\mu} \left( f_{L,q}P_L +f_{R,q}P_R \right)q,
\end{equation}
where $P_{L,R}$ are the projectors $P_L=(1-\gamma_5)/2$ and $P_R=(1+\gamma_5)/2$. We use a supercript ``$0$'' to denote the tree level standard model value of the coupling,
\begin{equation}
f_{L,q}^0=t^3_q -{\rm sin}^2\theta_W Q_q \qquad f_{R,q}^0=-{\rm sin}^2\theta_W Q_q.
\end{equation}
We take the value of ${\rm sin}^2\theta_W$ from the measured vector and axial vector lepton couplings at the $Z$ pole; then corrections to $Z$ vacuum polarization effects from the octet scalars are absorbed into it.

  \begin{fmffile}{ZbbGraphs} % requires ZbbGraph* files
\newenvironment{Zbb} 
   {\begin{fmfgraph*}(35,25)
      \fmfleft{Z}\fmfright{bb,b}
      \fmflabel{$\bar{b}$}{bb}
      \fmflabel{$Z$}{Z}\fmflabel{$b$}{b}}
   {\end{fmfgraph*}}

\begin{figure}
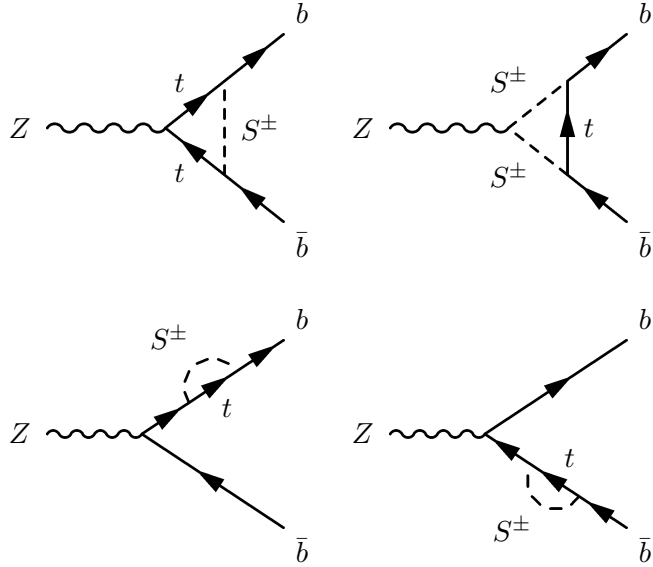

\centering
\begin{Zbb}
   \fmf{boson}{Z,Zv}
   \fmf{fermion}{bb,fbv}
   \fmf{fermion,label=$t$,label.side=left}{fbv,Zv}
   \fmf{fermion, label=$t$, label.side=left}{Zv,fv}
   \fmf{fermion}{fv,b}
   \fmffreeze
   \fmf{dashes, label=$S^{\pm}$, label.side=right}{fbv,fv}
%\fmfdot{Zv,fbv,fv}
 \end{Zbb}
\hspace{8mm}
\begin{Zbb}
   \fmf{boson}{Z,Zv}
   \fmf{dashes, label=$S^{\pm}$,label.side=left}{fbv,Zv}
   \fmf{dashes,label=$S^{\pm}$,label.side=left}{Zv,fv}
   \fmf{fermion}{bb,fbv}\fmf{fermion}{fv,b}
   \fmffreeze
   \fmf{fermion,label=$t$,label.side=right}{fbv,fv}
%\fmfdot{Zv,fbv,fv}
 \end{Zbb}\\[2\baselineskip]
\begin{Zbb}
  \fmf{boson}{Z,v1}
  \fmf{fermion,tension=1/3}{bb,v1}
  \fmf{fermion}{v1,v2}
  \fmf{fermion,label=$t$,label.side=right}{v2,v3}
  \fmf{dashes,left,tension=0,label=$S^{\pm}$}{v2,v3}
  \fmf{fermion}{v3,b}  
\end{Zbb} \hspace{8mm}
\begin{Zbb}
  \fmf{boson}{Z,v1}
  \fmf{fermion}{bb,v3}
  \fmf{fermion,label=$t$,label.side=right}{v3,v2}
  \fmf{dashes,left,tension=0,label=$S^{\pm}$}{v3,v2}
  \fmf{fermion}{v2,v1}  
  \fmf{fermion,tension=1/3}{v1,b}
\end{Zbb} 
\caption{Feynman diagrams contributing to $Z b \bar{b}$ vertex correction.}\label{ZbbFeynGraphs}
\end{figure}
\end{fmffile}

The down Yukawa coupling parameter, $\eta_D$, is constrained by the $B \rightarrow X_s \gamma$ partial width to be significantly less than $m_t / m_b$ when $\eta_U \sim 1$ and $ m_S \lesssim 5 {\rm TeV}$. Therefore, we neglect the down Yukawa coupling to the $b$ quark, as well as neglecting quark masses other than the top. Then the octet scalars give a one-loop correction to the left-handed bottom coupling from the vertex diagrams and bottom quark wave function renormalization diagrams pictured in Fig.~\ref{ZbbFeynGraphs}. Writing,
\begin{equation}
f_{L,b}=f_{L,b}^{\rm SM}+\delta f_{L,b},
\end{equation}
it is conventient to decompose the correction to the left-handed bottom quark coupling as,
\begin{equation}
\label{deltaf}
\delta f_{L,b}= -{4 \over 3}\left({1 \over 16 \pi^2}\right)|\eta_U|^2 |V_{tb}|^2 {m_t^2 \over v^2}\left[ f_{L,b}^0 {\cal A} +f_{R,t}^0 {\cal B}\right].
\end{equation}
We find that,
\begin{eqnarray}
{\cal A}&=& 2\int_0^1 {\rm d}x \int _0^{1-x}{\rm d}y\left[- {\rm ln}\left({m_t^{2} x+m_S^2(1-x) \over m_S^2(x+y)+m_t^2(1-x-y)-M_Z^2xy}\right) \right.\nonumber \\
&+&\left.{m_t^2 \over m_t^2(x+y)+m_S^2(1-x-y)-M_Z^2xy} \right]
\end{eqnarray}
and
\begin{eqnarray}
{\cal B}&=&1+2\int_0^1 {\rm d}x \int _0^{1-x}{\rm d}y\left[{\rm ln}\left({m_t^2(x+y)+m_S^2(1-x-y)-M_Z^2xy \over m_S^2(x+y)+m_t^2(1-x-y)-M_Z^2xy}\right)\right. \nonumber \\
&-&\left. {m_t^2/2+xyM_Z^2 \over m_t^2(x+y)+m_S^2(1-x-y)-M_Z^2xy} \right].
\end{eqnarray}
Here $m_S$ is the mass of the charged octet scalars. Note that in the limit $m_S>>m_t>>M_Z$,
\begin{equation}
{\cal A}=2{\cal B}=2{m_t^2 \over m_S^2}{\rm ln}\left(m_S^2 \over m_t^2 \right).
\end{equation}

We compute $\delta R_b$ using the formula,
\begin{equation}
\delta R_b \simeq 2 R_b^0(1- R_b^0) \left({f_{L,b}^0 \delta f_{L,b} \over (f_{L,b}^0)^2+(f_{R,b}^0)^2} \right)\simeq -0.78~\delta f_{L,b},
\end{equation}
where $\delta f_{L,b}$ comes from the vertex and $b$ quark field renormalization diagrams in Fig.~\ref{ZbbFeynGraphs} and is given above.

\begin{figure}
\centering
\includegraphics{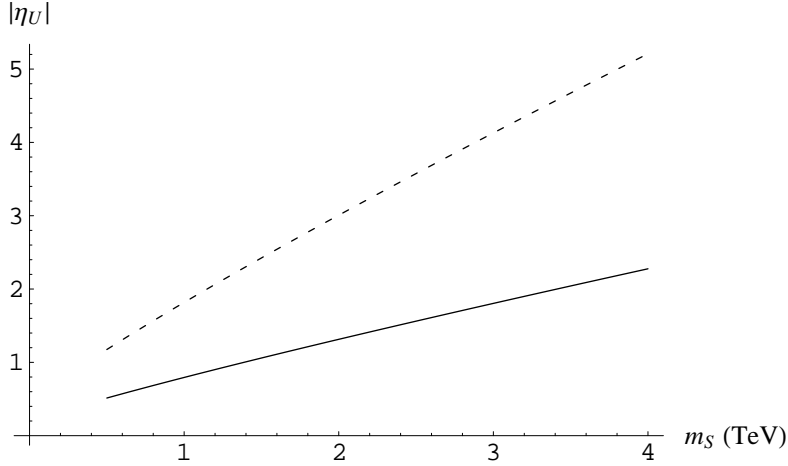}

\caption{ One (solid line) and two (dashed line) standard deviation exclusion contours due to $R_b$. (Parameter space above these lines is excluded.) The curves were calculated using $m_t = 170.9$ GeV, $\sin^2 \theta_{\text{eff, lept}} = 0.23153$, $m_Z = 91.1876$~GeV, and $v = 246$~GeV. }\label{RbEtaBoundPlot}
\end{figure}

According to the Particle Data Group~\cite{pdg}, the observed value for $R_b$ at the Z pole is 
\begin{equation}
R_b = 0.21629 \pm 0.00066  \qquad (\text{experiment})
\end{equation}
and the Standard Model predicted value is 
\begin{equation}
R_b = 0.21578 \pm 0.00010 \qquad (\text{SM prediction}). 
\end{equation}
Noting that the right-hand side of Eq.~(\ref{deltaf})  is positive, the following one sigma bound,
\begin{equation}
\delta f_b^L < 0.00020,
\end{equation}
on the contribution of the octet scalars to $\delta f_b^L$ follows from comparing the experimental value and the standard model prediction for $R_b$.

In Fig.~\ref{RbEtaBoundPlot} we plot the one and two sigma curves that bound the excluded $|\eta_U|$-$m_{S}$ parameter space. For example, if $m_{S} = 1 \text{TeV}$, then $|\eta_U|$ must be less than $0.8$ to agree with experiment to within $1 \sigma$, or less than $1.8$ to agree to within $2 \sigma$. 

\section{The cross section for single color octet scalar production via gluon fusion}

There are two neutral color octet scalars in the model we are considering. They are the scalar and pseudoscalar states destroyed by the real and imaginary parts of the field $S^{0A}$. The one loop gluon fusion rate~\cite{georgi} for producing these states singly is related to their two gluon decay rates by the standard formula,\footnote{See, for example, \cite{zerwas1995}.}
\begin{equation}
\sigma(p p \rightarrow S^0_{R,I}X)=\Gamma(S^0_{R,I}\rightarrow gg) \xi \left({16 \pi^2 \over s m_S}\right)\int_{m_S^2 / s}^1 \frac{dx}{x} g(x) g(m_S^2 / s x),
\end{equation}
 where $\xi=1/16$ is a spin-color factor that takes into account the interchange of summed-over and averaged-over states in the production cross section and decay rate.
% omitted: the fact that spins and colors are averaged over for the initial state and summed over the final state and % the initial states and final states are interchanged for the production cross section and the decay rate.
 
We begin by discussing the color octet scalar state's production. To simplify the analysis we assume that $CP$ nonconservation is small and take $\eta_U$, $\lambda_{4,5}$ to be real. Neglecting the mass differences between the various charged and neutral color octet scalar states we find that 

%\begin{eqnarray}
%\label {decayrate}
%&&\Gamma(S^0_R \rightarrow gg)=
%{G_F m_S^3 \alpha_s(m_S)^2 \over {\sqrt 2}~ 2^10 \pi^3 }\left( C_1 \eta_U^2 |I(m_t^2/m_S^2)|^2 - \right. \nonumber \\
%&& \left. 3C_2 \eta_U (\lambda_4+\lambda_5){v^2 \over m_S^2}\left({\pi^2 \over 9}-1\right) {\rm Re } I(m_t^2/m_S^2) +  {9 \over 4}C_3 (\lambda_4+\lambda_5)^2 {v^4 \over m_S^4}\left({\pi^2 \over 9}-1 \right)^2\right).
%\end{eqnarray}

\begin{multline}
\label {decayrate}
\Gamma(S^0_R \rightarrow gg)=
{G_F m_S^3 \alpha_s(m_S)^2 \over {\sqrt 2}~ 2^{10} \pi^3 } \Biggl[ C_1 \eta_U^2 |I(m_t^2 / m_S^2)|^2 + \\
 3 C_2 \eta_U  (\lambda_4+\lambda_5){v^2 \over m_S^2} \bigl({\pi^2  \over  9} -1\bigr) {\rm Re}I(m_t^2 / m_S^2) + {9 \over 4} C_3 (\lambda_4+\lambda_5)^2 {v^4 \over m_S^4}\bigl({\pi^2 \over 9}-1\bigr)^2 \Biggr].
\end{multline}
%where $z = m_t^2 / m_S^2$.

In Eq.~(\ref{decayrate}), $I(z)$ is the familiar factor from standard model Higgs decay. Assuming $z<1/4$ it is given by,
\begin{equation}\label{higgsFactor}
I(z)=2z+z(4z-1) {f(z)  \over 2} \qquad \qquad f(z) = \left({\rm ln}\left({1+\sqrt{ 1-4z} \over  1-\sqrt{ 1-4z}}\right) - i \pi \right)^2.
\end{equation}
The factors $C_j$ in Eq.~(\ref{decayrate}) are the color factors,\footnote{The unique symmetric invariant, $d^{ABC}$, of $SU(n)$ for $n \geq 3$ is given by $d^{ABC} = 2 \tr(\{T^A,T^B\} T^C)$ where $T^A$ are the fundamental representation matrices. Recall that the structure constants, $f^{ABC}$, are given by $f^{ABC} = - 2 i \tr([T^A,T^B] T^C)$. }
\begin{equation}
C_1=\sum (d^{ABC})^2 ={40 \over 3},~~~ C_2= \sum d^{ABC}d^{GFC}f^{AEF}f^{BGE}=-20,
\end{equation}
and
\begin{equation}
C_3=\sum (d^{GFC}f^{AEF}f^{BGE})^2= 30.
\end{equation}
The last two terms in~\eqref{decayrate}, which come from the octet scalar loops, are much smaller\footnote{This is true for $\lambda_{4,5} \sim 1$.} than the top loop contribution. (See Fig.~\ref{Sproduction}.) This is partly due to the factor of $(\pi^2/9-1)$.  For the pseudoscalar we find,
\begin{equation}
\Gamma(S^0_I \rightarrow gg) = {G_F \alpha_s(m_S)^2 m_t^4\over {m_S \sqrt 2}~ 2^{12} \pi^3 } C_1 \eta_U^2 \left\vert  f(m_t^2 / m_S^2) \right\vert^2 ,
\end{equation}
where $f(z)$ is as in~\eqref{higgsFactor}.
The pseudoscalar rate is due solely to a top loop and is related to that of a heavy color singlet by a simple multiplicative factor.

  \begin{fmffile}{sProdGraphs}

\newenvironment{ggS}
   {\begin{fmfgraph*}(40,25)
      \fmfleft{g1,g2}\fmfright{S}
      \fmflabel{$g$}{g1}
      \fmflabel{$g$}{g2}
      \fmflabel{$S^0_R$}{S}
      \fmf{gluon}{g1,v1}
      \fmf{gluon}{g2,v2}
      \fmf{scalar}{v3,S}}
   {\end{fmfgraph*}}

\begin{figure}
\centering
\begin{ggS}
  \fmf{quark,label=$t$,label.side=right}{v3,v2}
  \fmf{quark}{v1,v3}
  \fmffreeze
  \fmf{quark}{v2,v1}
\end{ggS}
\hspace{8mm} 
\begin{ggS}
  \fmf{scalar,label=$S^{\pm ,, 0}$,label.side=right}{v3,v2}
  \fmf{scalar}{v1,v3}
  \fmffreeze
  \fmf{scalar}{v2,v1}
\end{ggS}
\hspace{8mm}
\begin{fmfgraph*}(40,25)
  \fmfleft{g1,g3,g2}\fmfright{S}
  \fmflabel{$g$}{g1}\fmflabel{$g$}{g2}
  \fmflabel{$S^0_R$}{S}
  \fmf{phantom}{g3,v1}
  \fmf{scalar}{v2,S}
  \fmf{scalar,left,label=$S^{\pm ,, 0}$}{v1,v2,v1}
  \fmffreeze
  \fmf{gluon}{g1,v1}
  \fmf{gluon}{g2,v1}
\end{fmfgraph*}
\caption{Diagrams contributing to $S^0_R$ production via gluon fusion. For real $\lambda_{4,5}$, only the top loop contributes to $S^0_I$ production.}\label{Sproduction}
\end{figure}
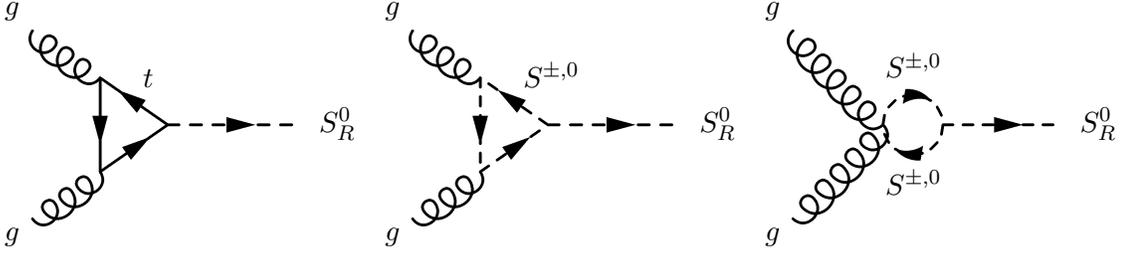
\end{fmffile}

Allowing CP violation mixes the scalar and pseudoscalar states. For example, the induced $g g S^0_R$ coupling acquires an axial contribution (and the $g g S^0_I$ coupling a non-axial contribution) proportional to ${\rm Im}(\eta_U)$. Also, the $S^0_I$ decay rate picks up a neutral scalar loop contribution proportional to ${\rm Im}(\lambda_4 + \lambda_5)$.

\begin{figure}
\centering
\includegraphics{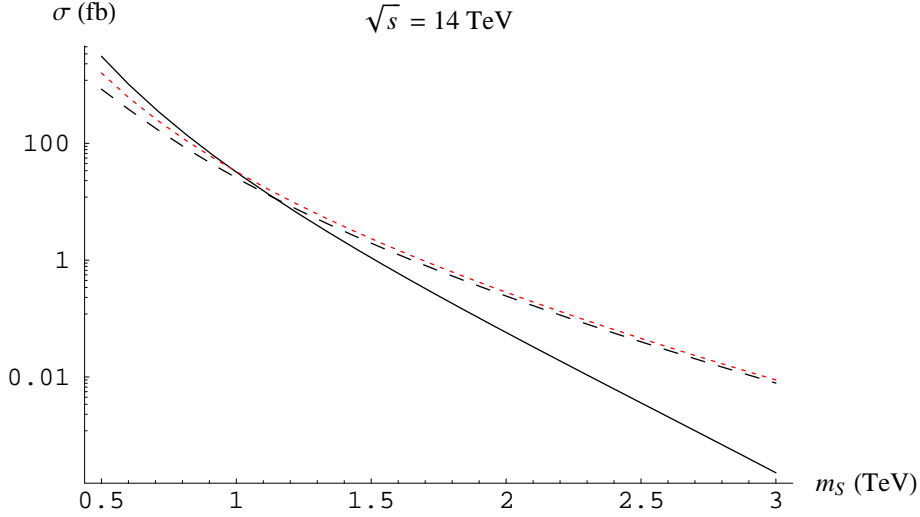}

\caption{ Production cross sections (in femto-barns) at LHC center of mass energy $\sqrt{s} = 14$ TeV for two real neutral scalars (solid line: $pp \rightarrow S^0_{R(I)} S^0_{R(I)} X$), one real neutral scalar (long dash: $pp \rightarrow S^0_{R} X$), and one real neutral pseudoscalar (short red dash: $pp \rightarrow S^0_{I} X$). The single color octet production cross sections are plotted with $\eta_U = 1$ and include only the top loop contribution, as the scalar loop contributions are negligible for $\lambda_{4, 5} \sim 1$. We used CTEQ5 next-to-leading order parton distribution functions \cite{cteq}, and we used the two-loop $\beta$ function to run $\alpha_s(m_Z) = 0.1216$ up to $\alpha_s(2 m_S)$ for scalar pair production and $\alpha_s(m_S)$ for single scalar production. The curves were calculated using $m_t = 170.9$~GeV. }\label{crossLHC}
\end{figure}

In Fig.~\ref{crossLHC} we plot the cross section for single $S^0_R$ and $S^0_I$ production, and real neutral scalar pair production as a function of $m_S$, using the above results and the pair production result from~\cite{wise2006a}.  For this plot the $\lambda_{4,5}$ terms are neglected and $\eta_U$ is set equal to unity. We expect that, just like in standard model Higgs production, the higher order QCD corrections are significant~\cite{nlo},~\cite{nnlo},~\cite{resum}. At a TeV, the single production rates begin to dominate over pair production.
%\footnote{Recall that below a TeV, the $R_b$ constraint favors $\eta_U < 1$, anyway.}

\begin{figure}
\centering
\includegraphics{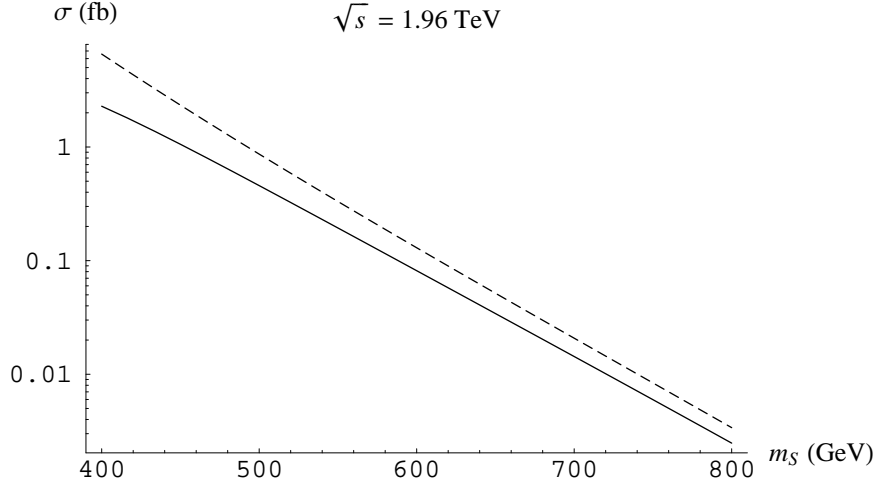}
\caption{
Production cross sections (in femto-barns) at Tevaton center of mass energy $\sqrt{s} = 1.96$~TeV for one real neutral scalar (solid line: $p \bar{p} \rightarrow S^0_{R} X$), and one real neutral pseudoscalar (dashed line: $p \bar{p} \rightarrow S^0_{I} X$). For this plot, $\eta_U = 1$ and the $\lambda_{4,5}$ terms in~\eqref{decayrate} are neglected. We used CTEQ5 next-to-leading order parton distribution functions \cite{cteq}, and we used the two-loop $\beta$ function to run $\alpha_s(m_Z) = 0.1216$ up to $\alpha_s(m_S)$. The curves were calculated using $m_t = 170.9$~GeV.
}\label{crossTEVA}
\end{figure}

In Fig.~\ref{crossTEVA} we plot the one-loop single production cross sections for $S^0_R$ and $S^0_I$ at the Tevatron. The tree level cross section for pair production is more than an order of magnitude smaller than the single production values in this mass and energy regime. For example, at $m_S = 500$ GeV, $\sigma(p \bar{p} \rightarrow S^0_{R(I)}S^0_{R(I)} X) = 10^{-3}~{\rm fb}$, while $\sigma(p \bar{p} \rightarrow S^0_{R} X) = 0.5~{\rm fb}$ and $\sigma(p \bar{p} \rightarrow S^0_{I} X) = 0.9~{\rm fb}$. 

In a recent paper~\cite{cdf}, the CDF collaboration presented preliminary limits from Run II on the production cross section for a ``$Z'$-like'' heavy neutral boson times its branching ratio to $t \bar{t}$ pairs. The analysis assumes that the boson appears as a Lorentzian enhancement in a limited region of the $M_{t \bar{t}}$ spectrum. Their preliminary limits do not exclude neutral octet scalars that decay mostly to $\bar t t$. %At 95\% confidence level, $Z'$-like bosons with mass $500$ GeV and $\sigma \dot B$ greater than $\sim 2.7$ pb 

\section {Concluding Remarks}
Minimal flavor violation is a convenient way to suppress flavor changing neutral currents when new degrees of freedom at the weak scale couple to quarks. If there are new scalar resonances with masses at the {\rm TeV} scale that couple to quarks, then minimal flavor violation implies that they are either color singlets or color octets with the same weak quantum numbers as the Higgs doublet. Models with two (or more) Higgs doublets have been studied extensively. The phenomenology of modes with an additional color octet scalar were studied in \cite{wise2006a}. Here we extended this work, calculating the constraint  on the strength of the Higgs coupling to up-type quarks that arises from precision electroweak data on $R_b$. We also computed the rate for single octet scalar production through gluon fusion. For color octet scalars with masses greater than $1~ {\rm TeV}$ this one-loop process can dominate over tree level pair production at $\sqrt{s} = 14$ TeV because the gluon parton distribution function increases rapidly as the momentum fraction decreases.
\begin{acknowledgments}
This work was supported in part by DOE grant number DE-FG03-92ER40701. M.G.~is supported in part by a National Defense Science and Engineering Graduate fellowship.
\end{acknowledgments}


\begin{thebibliography}{99}

 \bibitem{mfv}
   R.~S.~Chivukula and H.~Georgi,
  %``Composite Technicolor Standard Model,''
   Phys.\ Lett.\  B {\bf 188}, 99 (1987);
  %%CITATION = PHLTA,B188,99;%%
   L.~J.~Hall and L.~Randall,
   Phys.\ Rev.\ Lett.\ {\bf 65}, 2939 (1990);
  G.~D'Ambrosio, G.~F.~Giudice, G.~Isidori and A.~Strumia,
  %``Minimal flavour violation: An effective field theory approach,''
  Nucl.\ Phys.\  B {\bf 645}, 155 (2002)
  [arXiv:hep-ph/0207036].
  %%CITATION = NUPHA,B645,155;%%
  
 \bibitem{wise2006a}
  A.~V.~Manohar and M.~B.~Wise,
  %``Flavor changing neutral currents, an extended scalar sector, and the  Higgs
  %production rate at the LHC,''
  Phys.\ Rev.\  D {\bf 74}, 035009 (2006)
  [arXiv:hep-ph/0606172].
  
  %\cite{Popov:2005wz}
\bibitem{popov}
  P.~Y.~Popov, A.~V.~Povarov and A.~D.~Smirnov,
  %``Fermionic decays of scalar leptoquarks and scalar gluons in the minimal
  %four color symmetry model,''
  Mod.\ Phys.\ Lett.\  A {\bf 20}, 3003 (2005)
  [arXiv:hep-ph/0511149].
  %%CITATION = MPLAE,A20,3003;%%
  
  %\cite{Manohar:2006gz}
\bibitem{wise2006b}
  A.~V.~Manohar and M.~B.~Wise,
  %``Modifications to the properties of a light Higgs boson,''
  Phys.\ Lett.\  B {\bf 636}, 107 (2006)
  [arXiv:hep-ph/0601212].
  %%CITATION = PHLTA,B636,107;%%

\bibitem{pdg}
	W.-M. Yao et al.~(Particle Data Group), J.\ Phys.\ G {\bf 33}, 1 (2006). 
	%Particle Data Group
	
%\cite{Georgi:1977gs}
\bibitem{georgi}
  H.~M.~Georgi, S.~L.~Glashow, M.~E.~Machacek and D.~V.~Nanopoulos,
  %``Higgs Bosons From Two Gluon Annihilation In Proton Proton Collisions,''
  Phys.\ Rev.\ Lett.\  {\bf 40}, 692 (1978).
  %%CITATION = PRLTA,40,692;%%	
	
\bibitem{zerwas1995}
  M.~Spira, A.~Djouadi, D.~Graudenz and P.~M.~Zerwas,
  %``Higgs boson production at the LHC,''
  Nucl.\ Phys.\  B {\bf 453}, 17 (1995)
  [arXiv:hep-ph/9504378].
  %%CITATION = NUPHA,B453,17;%%
  
\bibitem{cteq}
 CTEQ Collaboration, [http://www.phys.psu.edu/$\sim$cteq].

\bibitem{nlo}
S.~Dawson,
  %``Radiative Corrections To Higgs Boson Production,''
  Nucl.\ Phys.\ B {\bf 359},  283 (1991);
  %%CITATION = NUPHA,B359,283;%%
  A.~Djouadi, M.~Spira and P.~M.~Zerwas,
  %``Production of Higgs bosons in proton colliders: QCD corrections,''
  Phys.\ Lett.\ B {\bf 264}, 440 (1991);
  %%CITATION = PHLTA,B264,440;%%
  C.~J.~Glosser and C.~R.~Schmidt,
  %``Next-to-leading corrections to the Higgs boson transverse momentum spectrum
  %in gluon fusion,''
  JHEP {\bf 0212}, 016 (2002);
  V.~Ravindran, J.~Smith and W.~L.~Van Neerven,
  %``Next-to-leading order QCD corrections to differential distributions of
  %Higgs boson production in hadron hadron collisions,''
  Nucl.\ Phys.\ B {\bf 634}, 247 (2002);
D.~de Florian, M.~Grazzini and Z.~Kunszt,
  %``Higgs production with large transverse momentum in hadronic collisions  at
  %next-to-leading order,''
  Phys.\ Rev.\ Lett.\  {\bf 82},  5209 (1999).

\bibitem{nnlo}
  R.~V.~Harlander and W.~B.~Kilgore,
  %``Next-to-next-to-leading order Higgs production at hadron colliders,''
  Phys.\ Rev.\ Lett.\  {\bf 88}, 201801 (2002);
%  [arXiv:hep-ph/0201206];
  %%CITATION = HEP-PH 0201206;%%
  C.~Anastasiou and K.~Melnikov,
  %``Higgs boson production at hadron colliders in NNLO QCD,''
  Nucl.\ Phys.\ B {\bf 646},  220 (2002);
%  [arXiv:hep-ph/0207004].
  %%CITATION = HEP-PH 0207004;%%
V.~Ravindran, J.~Smith and W.~van Neerven,
Nucl.\ Phys.\ B {\bf 665},  325 (2003).

\bibitem{resum}
S.~Catani, D.~de Florian, M.~Grazzini and P.~Nason,
  %``Soft-gluon resummation for Higgs boson production at hadron colliders,''
 JHEP {\bf 0307},  028 (2003);
G.~Bozzi, S.~Catani, D.~de Florian and M.~Grazzini,
  %``The q(T) spectrum of the Higgs boson at the LHC in QCD perturbation
  %theory,''
  Phys.\ Lett.\ B {\bf 564}, 65 (2003);
G.~Bozzi, S.~Catani, D.~de Florian and M.~Grazzini,
  %``Transverse-momentum resummation and the spectrum of the Higgs boson at the
  %LHC,''
  Nucl.\ Phys.\ B {\bf 737}, 73 (2006).
  %%CITATION = HEP-PH 0508068;%%

\bibitem{cdf} Kagan et al.\ (CDF),  ``Limit on Resonant $t \bar{t}$ Production in $p \bar{p}$ Collisions at $\sqrt{s} = 1.96$ TeV,'' CDF Note 8675 (2007) [http://www-cdf.fnal.gov]

\end{thebibliography}
\end{document}